\documentclass[aps,prl,superscriptaddress,twocolumn]{revtex4-1}
\usepackage{graphics}
\usepackage{graphicx}
\usepackage{color}
\usepackage{amsmath, amsthm, amssymb}
\usepackage{subfigure}
\usepackage{ulem}
\usepackage{natbib}

\begin{document}

\title{Spin-based Mach-Zehnder interferometry in topological insulator p-n junctions}
\author{Roni Ilan}
\email{ rilan@berkeley.edu}
\affiliation{Department of Physics, University of California, Berkeley, California 94720, USA}

\author{Fernando de Juan}
\affiliation{Department of Physics, University of California, Berkeley, California 94720, USA}
\affiliation{Materials Sciences Division, Lawrence Berkeley National Laboratory, Berkeley, CA 94720}

\author{Joel E. Moore}
\affiliation{Department of Physics, University of California, Berkeley, California 94720, USA}
\affiliation{Materials Sciences Division, Lawrence Berkeley National Laboratory, Berkeley, CA 94720}


\begin{abstract}
A p-n junction, an interface between two regions of a material populated with carriers of
opposite charge, is a basic building block of solid state electronic devices. From the fundamental physics perspective, it often serves as a tool to reveal the unconventional transport behavior of novel materials. In this work, we show that a p-n junction made from a three dimensional topological insulator (3DTI) in a magnetic field realizes an electronic Mach-Zehnder interferometer with virtually perfect visibility. This is owed to the confinement of the topological Dirac fermion state to a closed two-dimensional surface, which offers the unprecedented possibility of utilizing external fields to design networks of chiral modes wrapping around the bulk in closed trajectories, without the need of complex constrictions or etching. Remarkably, this junction also acts as a spin filter, where the path of the particle is tied to the direction of spin propagation. It therefore constitutes a novel and highly tunable spintronic device where spin polarized input and output currents are naturally formed and could be accessed and manipulated seperately.
\end{abstract}

\maketitle

Three dimensional topological insulators were discovered shortly after their prediction, by photoemission experiments demonstrating the existence of a single Dirac cone in their energy spectrum~\cite{RevModPhys.82.3045,Review}. In transport, the study of their properties has proved to be very challenging, mainly due to bulk carriers obscuring effects coming from the surface states, and limited control over chemical potentials. Recently, remarkable progress has been made both in growth of new types of topological insulator materials as well as measuring techniques: a new generation of compounds and alloys exhibit a low contribution to transport coming from bulk carriers, and improvements have been made in gating methods~\cite{zhang2011band,kong2011ambipolar,arakane2012tunable,AndoReview}. Hence, efforts can now be focused more on exploring exotic phenomena. 

Dirac fermions confined to the surface of topological insulators are expected to exhibit remarkable effects: Chiral modes, topologically protected perfectly transmitting modes, as well as perfect Andreev reflection, are only some of the phenomena predicted to appear in transport~\cite{qi2011topological,L09,ZWX12,SRA12,bardarson2013quantum,PhysRevLett.100.096407,PhysRevLett.102.216403,PhysRevLett.102.216404}. In the presence of strong magnetic fields, Shubnikov de Haas oscillations and a quantum Hall effect typical to Dirac particles have already been demonstrated~\cite{PhysRevLett.106.126803,cheng2010landau,qu2010quantum,analytis2010two,kozlov2014transport,HallTopBottom,2014arXiv1409.3778X}. However, features unique to topological insulators such as the spin structure of the quantum Hall edge modes are yet to be observed.  Here, we predict that a simple and rather common setup could pin-point these properties in the form of an interferometric measurement in a device made only from a film of TI, a top gate, and an external magnetic field. We show that a thin film of a topological insulator in the quantum Hall regime, partly covered by a gate, forms a p-n junction which in the presence of an in-plane flux functions as a condensed matter realization of a Mach-Zehnder interferometer. The oscillations in the conductance of this device are directly linked to the spin texture of the quantum Hall chiral modes, a manifestation of the spin-momentum locking property of TI surface states.  

In addition to providing a testbed to reveal the unique transport properties of 3DTI, the structure we present here makes a spintronic device with a very high level of spin control. The two-path interferometer takes as inputs one dimensional currents that are naturally spin polarized without the aid of ferromagnetic leads, and generates as outputs two spin polarized channels that are spatially separated. The interference loop is controllable with the application of an in-plane flux, which determines how much spin current is carried in each output channel. In a two terminal geometry, the behavior of the conductance of this device is similar to that of the spin-FET, proposed by Datta and Das over two decades ago \cite{DD90}. But our device functions more generally as a spin filter, granting separate access to the different spin orientations produced as outputs, and allowing to tune their relative amplitudes.  

The idea of using a p-n junction to look for novel physical effects in Dirac materials traces back to the remarkable experiments done in graphene~\cite{WDM07,SHG09,YK09}, where the interplay between Klein tunneling and the Quantum Hall Effect is responsible for characteristic transport signatures evidencing Dirac fermions~\cite{AL07,tworzydlo2007valley}. A p-n junction has also been studied theoretically before in TI~\cite{PhysRevB.85.235131}, yet the effect we predict here was overlooked. When designing transport devices made out of a film of a 3DTI, a single flat surface is generally assumed for simplicity, while the effect described here requires considering the full 2D surface of a 3D bulk material . While superficially similar to ``one quarter of'' graphene, the surface state of TI is fundamentally different in that it lives on a closed surface with no edges, embedded in three dimensional space. The usual edge state picture of the Quantum Hall Effect is thus replaced
by the more general principle that chiral modes appear between regions with different local Hall conductance. This is the main reason our design is unique to topological insulators and cannot be implemented in planar geometries. 
\begin{figure*}[t]
\begin{center}
\includegraphics[width=18cm]{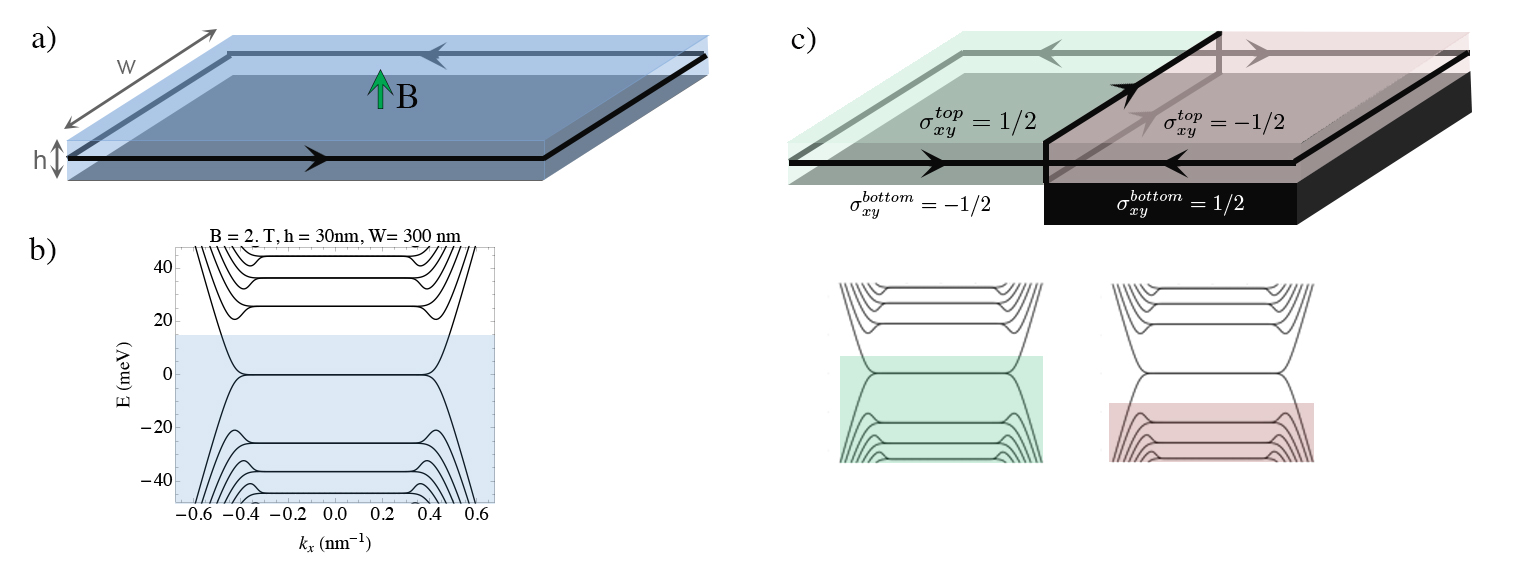}
\caption{\textbf{A p-n junction from a three dimensional topological insulator film.} a) A film of a 3DTI of height $h$ and width $W$ is placed under a strong perpendicular magnetic field, such that the top and bottom surfaces are gapped, and the side surfaces carry chiral modes. b) The energy spectrum of the surface states for an infinite film with $B=2$ T, $h=30$ nm, $W=300$ nm. The blue region marks the chemical potential, such that the system is in the lowest Landau level and there is a single mode running along the side surfaces. c) A p-n junction in a magnetic field: a bottom gate shifts the chemical potential of the right side of the sample below the Dirac point, such that $\sigma_{xy}$ for the top and the bottom surfaces in this part of the film are reversed. The film is then traversed by two additional chiral modes forming a closed loop around the bulk. }
\label{fig1}
\end{center}
\end{figure*}

\section{results}

Consider a thin film of a 3DTI in the presence of a strong perpendicular magnetic field, as shown in Fig. \ref{fig1}(a), so that the the top and the bottom surfaces are in the Quantum Hall regime with well developed Landau Levels [see Fig.~\ref{fig1}(b)]. Since the surface state is a single Dirac fermion, when the $n = 0$ Landau Level is filled, the top surface has Hall conductance $\sigma_{xy} = 1/2$ in units of $e^2/h$. The bottom surface, penetrated by a magnetic field opposite in direction with respect to the surface normal, has $\sigma_{xy}=-1/2$ \cite{qi2011topological}. This implies that the side surfaces, which are interfaces between regions where $\sigma_{xy}$ changes by one unit, will carry a single chiral mode, similar to the one appearing in a quantum Hall state with $\sigma_{xy} = 1$ in a strictly 2D system.

To form a p-n junction, imagine gating half of the sample so that the chemical potential in the gated region resides between the $n=0$ and $n=-1$ Landau levels. This flips the signs of the Hall conductances of the top and bottom surfaces, and as a result, the chirality of the current flowing on the side surfaces in the gated region is reversed. At the p-n junction, currents of opposite chirality meet at a vertex $v1$ on one side surface, and part from a vertex $v2$ on the other side. In addition, since the Hall conductance now changes sign not only from top to bottom, but also from right to left, two more chiral modes co-propagating along the junction connect $v_1$ and $v2$, one running on the top and one on the bottom surface. Therefore, the p-n junction realizes a network of chiral modes with the structure of an interferometer, as shown in fig. Fig.~\ref{fig1}(c). Crucially, because the modes traversing the film and forming the loop live in different surfaces, the loop encloses a finite area and an in-plane magnetic field can be used to thread flux through it, allowing the study of interference effects.
\begin{figure*}[t]
\begin{center}
\includegraphics[width=16cm]{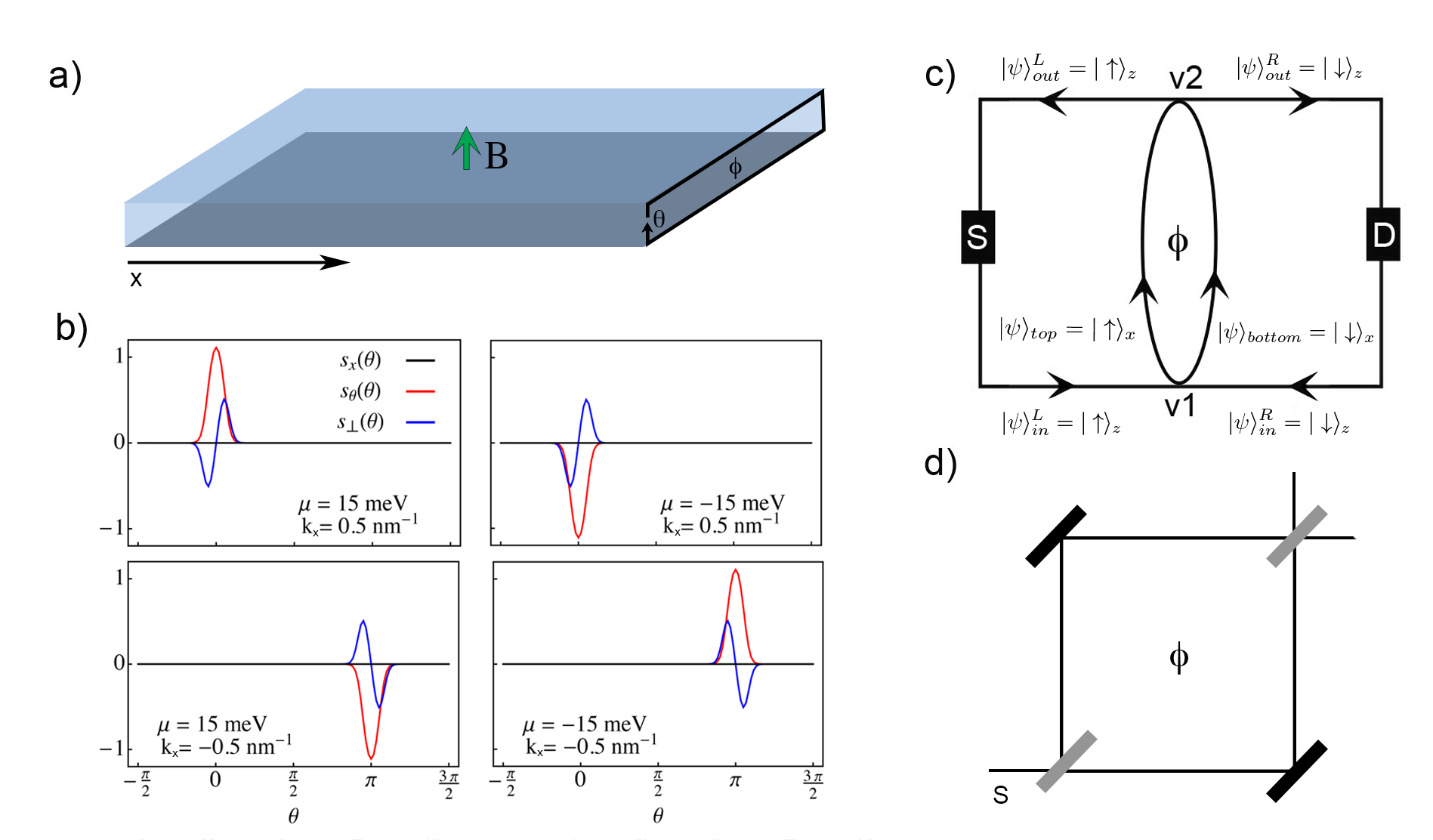}
\caption{\textbf{Spin textures and the p-n junction as a spin based Mach Zehnder interferometer}. a) A thin film of a 3DTI in the presence of a perpendicular and in-plane magnetic field. The in-plane magnetic field is enclosed as flux through the interference loop. b) The spin components of the front and back side surfaces. $\theta$ is the coordinate running along the circumference of the film as indicated in (a), Spin vectors are described in the tangent basis $\vec s = (s_x,s_\theta,s_\perp)$ which rotates along $\theta$, keeping $s_\perp$ aligned with the surface normal. c) A schematic of the interferometer and the effective spin states on each arm. The vertices on the side surfaces labeled $v1,v2$ scatter the incoming spin states into rotated spin states on the top and bottom surfaces encircling the flux. d) The original optical Mach Zehnder interferometer. The gray rectangles are the beam splitters. }
\label{fig2}
\end{center}
\end{figure*}

The key property of this network is that the chiral modes forming it have characteristic spin polarizations due to spin-momentum locking. To see this explicitly, we model our system as a Dirac fermion living on the surface of a rectangular slab placed in magnetic field~\cite{JIB14}, as in Fig.~\ref{fig2}(a), and obtain the wave functions numerically (see Methods section).  The surface is parametrized with two coordinates $(x,\theta)$, where $x$ runs along the slab (perpendicular to the p-n junction) and $\theta$ encircles the slab. Spin vectors are described using the basis $\vec s = (s_x,s_\theta,s_\perp)$ which rotates when moving along $\theta$, keeping $s_\perp$ aligned with the surface normal. The expectation value of the different spin components is plotted as a function of $\theta$ in Fig.~\ref{fig2}(b). As the figure shows, the average spin density on the side surfaces amounts to a net polarization perpendicular to the direction of the current flow. Changing the chirality of the current, either by flipping the direction of the field, or by changing the sign of the chemical potential, also reverses the direction of spin polarization. Hence, the chiral modes propagating along the side surfaces of the p-n junction carry opposite spin on opposite sides of the junction. Note that the effective spin always lies in the surface plane, and points from the region with $\sigma_{xy} = -1/2$ to the one with $\sigma_{xy} = 1/2$, and the same is correct for the modes traversing the junction.

The chiral modes traversing the sample on top and bottom surfaces have a different spin polarization than those at the side surfaces, and this is fundamental in generating interference effects in transport through the p-n junction. At the vertex $v1$, incoming modes have $\pm \theta$ polarization but outgoing modes have $\pm x$ polarization. Choosing specific spin operators for concreteness as $(\hat s_x, \hat s_\theta, \hat s_\perp) = (\sigma_x,\sigma_z,\sigma_y)$, an electron entering the vertex must be scattered into the outgoing modes according to the basis change
\begin{eqnarray}
|\uparrow\rangle_z=\frac{1}{\sqrt{2}}|\uparrow\rangle_x+\frac{1}{\sqrt{2}}|\downarrow\rangle_x\\
|\downarrow\rangle_z=\frac{1}{\sqrt{2}}|\uparrow\rangle_x-\frac{1}{\sqrt{2}}|\downarrow\rangle_x
\end{eqnarray}
which means that the vertex $v1$ acts as a beam splitter. The same occurs at the vertex $v2$, with $x$ and $z$ interchanged.  A schematic of this interferometer is shown in Fig.~\ref{fig2}(c). Therefore, our setup has the exact structure of a two path interferometer, sketched in Fig.~\ref{fig2}(d). The path of an electron emitted from a source contact at the left side splits into two paths at $v1$, propagating across the sample, and recombining at $v2$. Flux through the loop introduces a relative phase between the two arms of the interferometer. 	
When this phase is zero, the electron recombines at $v_2$ with its spin orientation unchanged and is perfectly reflected, while a phase of $\pi$ causes the incoming spin to be completely flipped so that perfect transmission occurs. 
Intermediate phases will result in an intermediate spin rotation and therefore, partial transmission. This spin-based interference effect can be directly observed in two-terminal transport.

\begin{figure*}[t]
\begin{center}
\includegraphics[width=8.5cm]{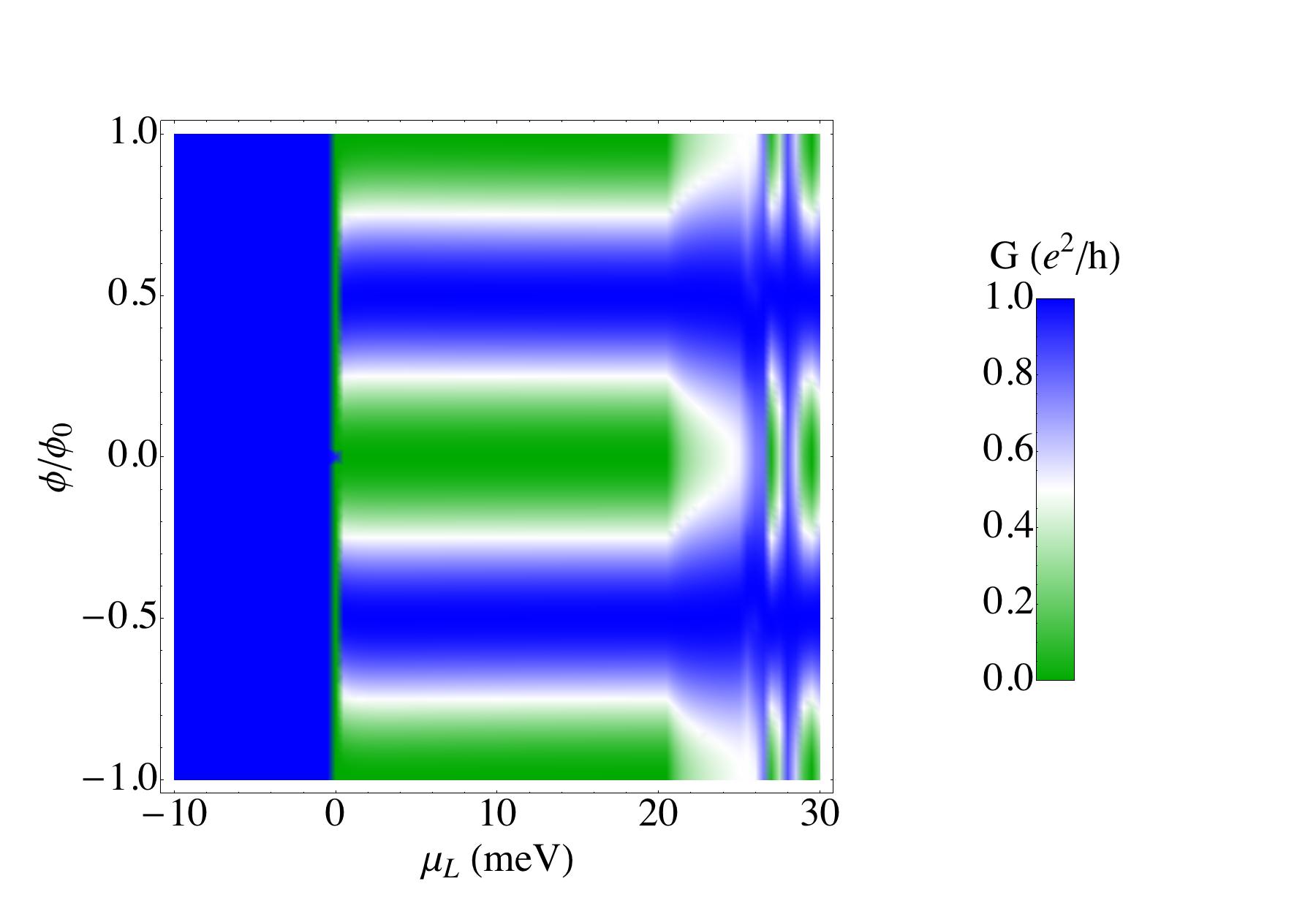}
\includegraphics[width=8.5cm]{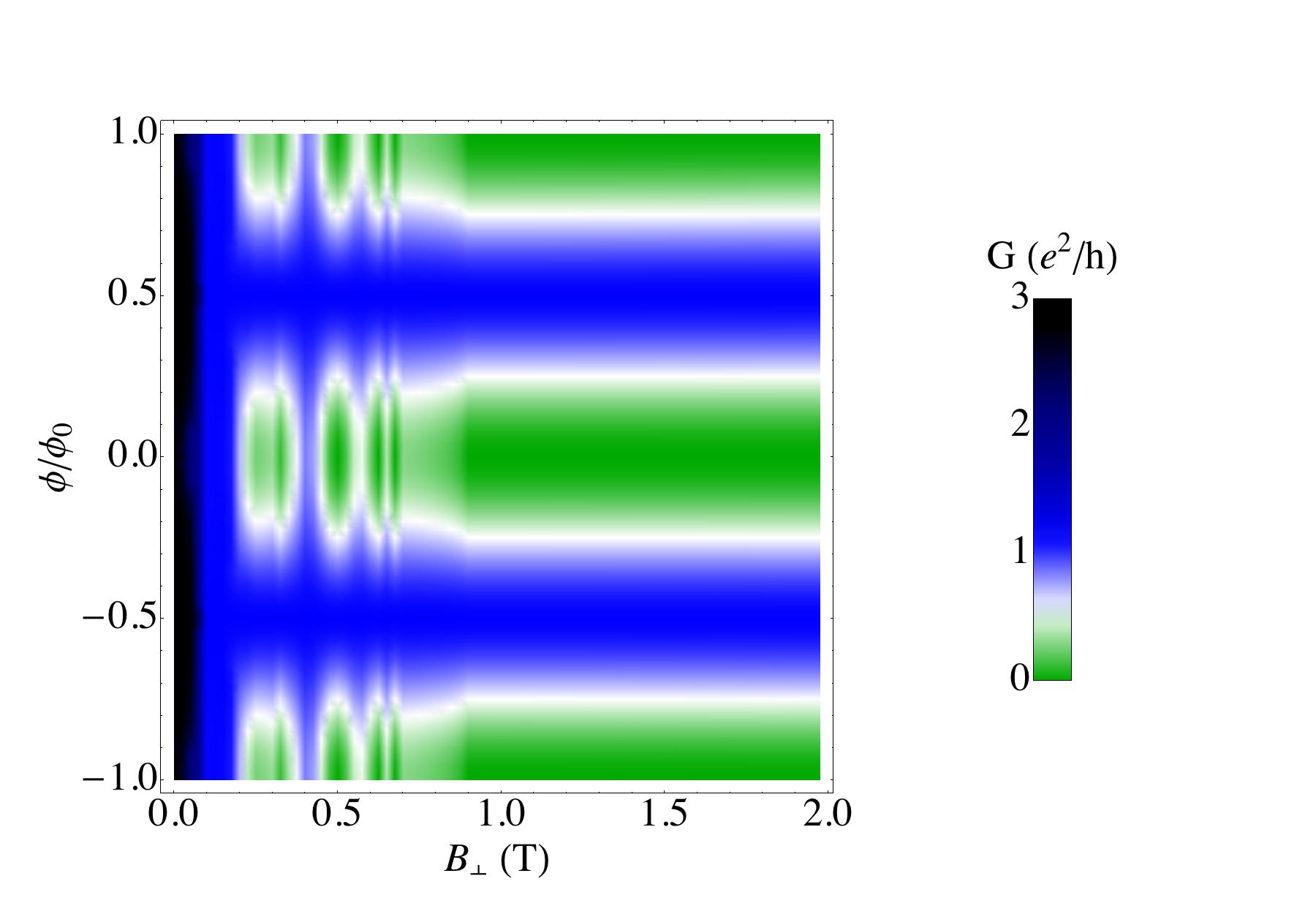}
\caption{\textbf{Conductance of a p-n junction as a function of flux, chemical potential, and magnetic field. }Left: Conductance as a function of $\phi$ and $\mu_L$, for a fixed $\mu_R=-15$ meV and $B_\perp = 2$ T. Right: Conductance as a function of $\phi$ and $B_\perp$ for fixed $\mu_R = -\mu_L = 15$ meV.  In both plots, once fully developed Landau levels are formed, and the chemical potential is in the lowest Landau level, the conductance follows equation~(\ref{eq:interference}).}
\label{2Dplots}
\end{center}
\end{figure*}

To make our above statements quantitative, we compute the two terminal conductance across the p-n junction from the scattering matrix $S$ of the device
\begin{equation}
S=\left(
\begin{array}{ccc}
  r  & t'  \\
  t  &r'      
\end{array}
\right)
\end{equation}
with $r,t, r',t'$ representing the reflection and transmission of incoming modes from a scattering region. The two terminal conductance from source to drain is simply calculated from the transmission matrix $t$ and is given by  
\begin{equation}\label{eq:conductance}
G=\frac{e^2}{h}|t|^2
\end{equation}
Denoting the two incoming modes as  $|\psi\rangle_{in}^L=|\uparrow\rangle_z$ and $ |\psi\rangle_{in}^R=|\downarrow\rangle_z$ (see Fig.~\ref{fig2})
the scattering matrix of the first vertex
 \begin{equation}
S_{v1}=1/\sqrt{2}(\sigma_x+\sigma_z)
\end{equation}
 accounts for the scattering of the incoming modes onto  $|\psi\rangle_{top}$ and $|\psi\rangle_{bottom}$, polarized along the $x$ direction.
The transfer matrix describing the propagation of the spinors $|\psi\rangle_{top}$ and $|\psi\rangle_{bottom}$ along the arms of the interferometer is diagonal, and contains two contributions. The first is a dynamical phase, $e^{ikW}$, where $k$  is the momentum of the particles and $W$ the width of the film. The second is a phase contributed by the flux enclosed in the loop, $\phi$. It is therefore given by 
\begin{equation}
T=e^{ikW+i\phi/2}e^{i\phi\sigma_z/2}
\end{equation}
Finally, from symmetry, the scattering matrix describing the second vertex  $v_2$ is $S_{v2}=S_{v1}$. Therefore the total scattering matrix for the interference loop is the product of all three matrices
\begin{equation}
S=S_{v1}TS_{v2}=
e^{ikW+i\phi/2}e^{i\phi\sigma_x/2}
\end{equation}
From this we find the simple expression
\begin{equation}\label{eq:interference}
G=\frac{e^2}{h}\sin^2\phi/2
\end{equation}
As a function of flux through the loop, the conductance oscillates between zero and $e^2/h$. 

We pause here to appreciate the fact that this interferometer has perfect visibility: the visibility of an interferometer is defined as $V=(I_{max}-I_{min})/(I_{max}+I_{min})$, where $I_{max(min)}$ are the maximum (minimum) of the current measured at the drain. Since the two beam splitters should split the current in half, we expect this interferometer to be very close to having unit visibility.

Equation~(\ref{eq:interference}) stems from the effective one dimensional model deduced from the spin densities we have calculated numerically. In order to confirm the validity of this result, we perform an exact numerical calculation of the conductance of the p-n junction on the slab, again using the scatting matrix approach. The Hamiltonian is solved in the left and right sides with chemical potentials $\mu_L>0$ and $\mu_R<0$ respectively, and the wavefunctions are matched at the interface to obtain the transmission matrix, from which the conductance is calculated in a similar fashion to equation~(\ref{eq:conductance}). Figure~\ref{2Dplots} shows the conductance as a function of flux through the loop and chemical potential or perpendicular field. As it shows, once Landau levels are well established and the two sides are brought into the lowest Landau level regime, the conductance oscillates with unit amplitude and a period of one flux quantum. This confirms the prediction of the effective model given by equation~(\ref{eq:interference}), and in particular the fact that the the interference pattern has unit visibility within the Dirac fermion model. We note that corrections may be introduced due to hexagonal warping~\cite{Fu09} and Zeeman coupling. The first becomes important at energies larger than $E^* \approx 230$ meV~\cite{Fu09}, and  the second affects states only below the Zeeman energy, which at $B_{\perp}=2$ T and $g$ as large as $20$ is only $E_Z = \mu_B g B_{\perp}/2\approx 1.1$ meV. Since we require only that the chemical potential must lie below the first Landau Level $E_{n=1} \approx 20-30$ meV, for a wide range of $\mu$ both effects can be safely neglected. 

\section{Discussion}
 
In summary, we have demonstrated that a p-n junction formed from a three dimensional topological insulator in the presence of a strong external perpendicular field and an in-plane flux can be turned into an interferometer which acts as a spin filter. The observation of conductance oscillations in this setup should serve as a clear-cut transport signature of spin momentum locking in 3DTI, and of the unusual spin texture of chiral modes in the quantum Hall regime. It is interesting to note the relation of our proposal with the electronic Mach-Zehnder interferometer realized in the beautiful experiment by Ji et al. \cite{MAchZehnder} using Quantum Hall edge states in a two dimensional electron gas. There, the effect seen is purely charge-based and lacks the spin-filter aspect discussed here, which is unique to TI. 

The oscillations in the two terminal conductance of our setup can also be thought of as an analog effect to the one appearing in the Datta-Das spin-FET \cite{DD90}.  There, the conductance between two ferromagnetic contacts can be changed by inducing a spin precession in the region that separates them through changing the spin-orbit coupling with a gate voltage. In our setup, the input channels are naturally spin polarized so that normal contacts can be used rather than ferromagnetic leads, and the spin precession is induced by the in-plane flux. Compared to the spin-FET, our device has the outstanding advantage of spin-filtering, by physically separating the different spin components on output, which could be accessed using an appropriate contact configuration.  

The authors are indebted to Yong P. Chen and Yang Xu for invaluable discussions. We also thank Ashvin Viswanath, Adolfo Grushin and Jens Bardarson for useful comments on the manuscript. The authors also acknowledge funding from DARPA FENA (R.I.), AFOSR MURI (F. d. J), NSF DMR-1206515 and Simons Foundation (J. E. M.). 
 
\section{Methods}
The Hamiltonian describing the Dirac fermion surface state of a 3DTI slab of height $h$ and width $W$ in a constant magnetic field is
\begin{equation}
H=-iv_F\hat s_\theta (\partial_x+ieA_x)+iv_F \hat s_x \frac{2\pi}{P}(\partial_\theta+i\frac{\phi}{2\pi})-\mu
\end{equation}
with $\theta \in [0,2\pi]$ the dimensionless coordinate that wraps around the slab (see Fig.~\ref{fig2}) and $x$ the coordinate along the slab, $P$ the perimeter $P=2h+2W$, $v_F$ the Fermi velocity of the particles, $\mu$ the chemical potential, $\phi$ the flux trough the film cross section in units of $2\pi \Phi_0$, with $\Phi_0=h/e$ the flux quantum, and $\hat s_i$ are the spin operators. The boundary conditions in the $\theta$ direction are anti-periodic \cite{bardarson2013quantum}. In this coordinate system, the magnetic field is $\pm B$ for the top and bottom surfaces and zero for the sides, and is reproduced with a vector potential $A_x=A_x(\theta)$ as the one given in ref.~\cite{JIB14}. The numerical diagonalization of this Hamiltonian can be done by expanding the eigenfunctions in angular momentum states $\psi_{k}(s) = \sum_n e^{i n \theta} \chi_{k,n}$ with $n$ half-integer, as described in detail in ref.~\cite{JIB14}. The energy spectrum obtained by that method is shown in Fig.~\ref{fig1} for $B=2 \rm{T}$, $h=30 \rm{nm}$, $W=300 \rm{nm}.$
 The spin polarization of the lowest Landau level side surface modes is obtained from the corresponding eigenstate $\psi_{k}(\theta)$ by calculating the expectation values $s_i(\theta)= \left< \psi_k^{\dagger}| \hat s_i |\psi_k \right>$, and is shown on Fig.~\ref{fig2}. 
 
A p-n junction is modeled by chemical potential that changes sign at $x=0$, $\mu(x) = \mu_L \theta(x) + \mu_R \theta(-x)$, with $\mu_L>0$ and $\mu_R<0$. To obtain the conductance of the junction, we compute its scattering matrix by matching the wavefunctions of the two sides at $x=0$. As in ref.~\cite{JIB14}, this calculation involves both propagating and evanescent modes at the Fermi level,  which can be computed using the transfer matrix approach \cite{LJ81,U97}. The relevant modes for the calculation are defined as follows. $\psi^{L-}_{\alpha}$ and $\psi^{L+}_{\alpha}$ with $\alpha = 1,\ldots, N_{\rm prop}$ and $\psi^{L,ev}_{\alpha'}$ with $\alpha' = 1, \ldots, N_{\rm ev}$ are incoming, outgoing, and evanescent modes at $x<0$. Similarly, for $x>0$ we have $\psi^{R-}_{\alpha}$, $\psi^{R+}_{\alpha}$,$\psi^{R,ev}_{\alpha'}$. In terms of these modes, the matching condition takes the form
\begin{align}
\psi^{L-}_{\alpha} + \sum_{\beta=1}^{N_{\rm prop,L}}r_{\alpha \beta}\psi^{L+}_{\beta} +
\sum_{\alpha'=1}^{N_{\rm ev,L}}
c_{\alpha \alpha'} \psi^{L,ev}_{\alpha'} \\
= \sum_{\beta=1}^{N_{\rm prop,R}}t_{\alpha \beta}\psi^{R+}_{\beta} +
\sum_{\alpha'=1}^{N_{\rm ev,R}}
c'_{\alpha \alpha'} \psi^{R,ev}_{\alpha'} 
\end{align}
where $r_{\alpha \beta}$ and $t_{\alpha \beta}$ are the elements of the reflection and transmission matrices. The conductance is given by $G = e^2/h {\rm tr}( t^{\dagger} t)$. Fig.~\ref{2Dplots} shows the conductance for different values of magnetic field, flux and chemical potentials.

\bibliography{chiral} 
\end{document}